\documentclass{article}

\usepackage{graphicx}
\usepackage{psfig}
\usepackage{epsfig}
\usepackage[round]{natbib}

\title{
Statistical modelling of tropical cyclone tracks: a semi-parametric
model for the mean trajectory
}

\begin{document}

\author{Tim Hall, GISS\footnote{\emph{Correspondence address}: Email: \texttt{tmh1@columbia.edu}}\\and\\
Stephen Jewson\\}

\maketitle

\begin{abstract}
We present a statistical model for the unconditional mean tracks of hurricanes.
Our model is a semi-parametric scheme that averages together observed hurricane
displacements. It has a single parameter that defines the averaging length scale,
and we derive the optimum value for this parameter using a jackknife.
The main purpose of this model is as a starting point for developing a
statistical model of hurricanes for use in the estimation of the
wind, rainfall and flooding risks.
The model also acts as an optimal filtering tool for estimating
mean hurricane tracks.
\end{abstract}

\section{Introduction}

We are interested in the question of how to estimate the wind, rainfall and flooding risks
caused by tropical cyclones. Various methods to evaluate these risks have been developed over
the last 20 years in industry and academia. The insurance industry, for instance,
uses such methods to inform the process of setting insurance rates.
Some of the models used in industry have been published, either in full or
in part (see, for example, \citet{drayton00} and \citet{clark86}).
The models developed by academics include those of
\citet{darling91},
\citet{chu98},
\citet{vickery00} and
\citet{emanuel05}.
The papers of~\citet{vickery00} and~\citet{emanuel05} contain a number of
references to other studies in this subject area.

The papers cited above use a range of methods for modelling tropical
cyclone risks. At present it seems that there has been no systematic
attempt to compare these different methods, and there is little
consensus as to which of the methods are the best. Our long-term
goal is to try and bring some more rigourous analysis to bear on
this question. However, our immediate goal is to start to develop a
hierarchy of models that simulate hurricane tracks.
The models in this hierarchy will be similar, at a
general level, to the models described by \citet{drayton00}, \citet{vickery00} and \citet{emanuel05}.

One of our specific aims is to build models using statistical
best-practice. There are various aspects to this. For instance, we
intend to test the models we develop using cross-validation (often known
in meteorology as `out-of-sample testing').
This restricts us to developing models that can be fitted to the observed
data automatically and efficiently (in a matter of minutes or hours), since
only such models can be tested in this way.
Also, we intend to follow a systematic modelling philosophy whereby
we start with a simple model and add complexity only to the extent
that it can be proven to improve the cross-validated results. The
model that we describe in this article is therefore rather simple,
as is appropriate for the first model in our hierarchy, but it sets the stage for
the development of more complex models that can easily be tested and
compared. By developing such a hierarchy we hope that we will
gradually converge on an accurate model that makes good use of the
available data but avoids overfitting. And maybe we will
learn something about tropical cyclones on the way, too.

\section{Data}

\begin{figure}[!htb]
  \begin{center}
    \scalebox{0.8}{\includegraphics{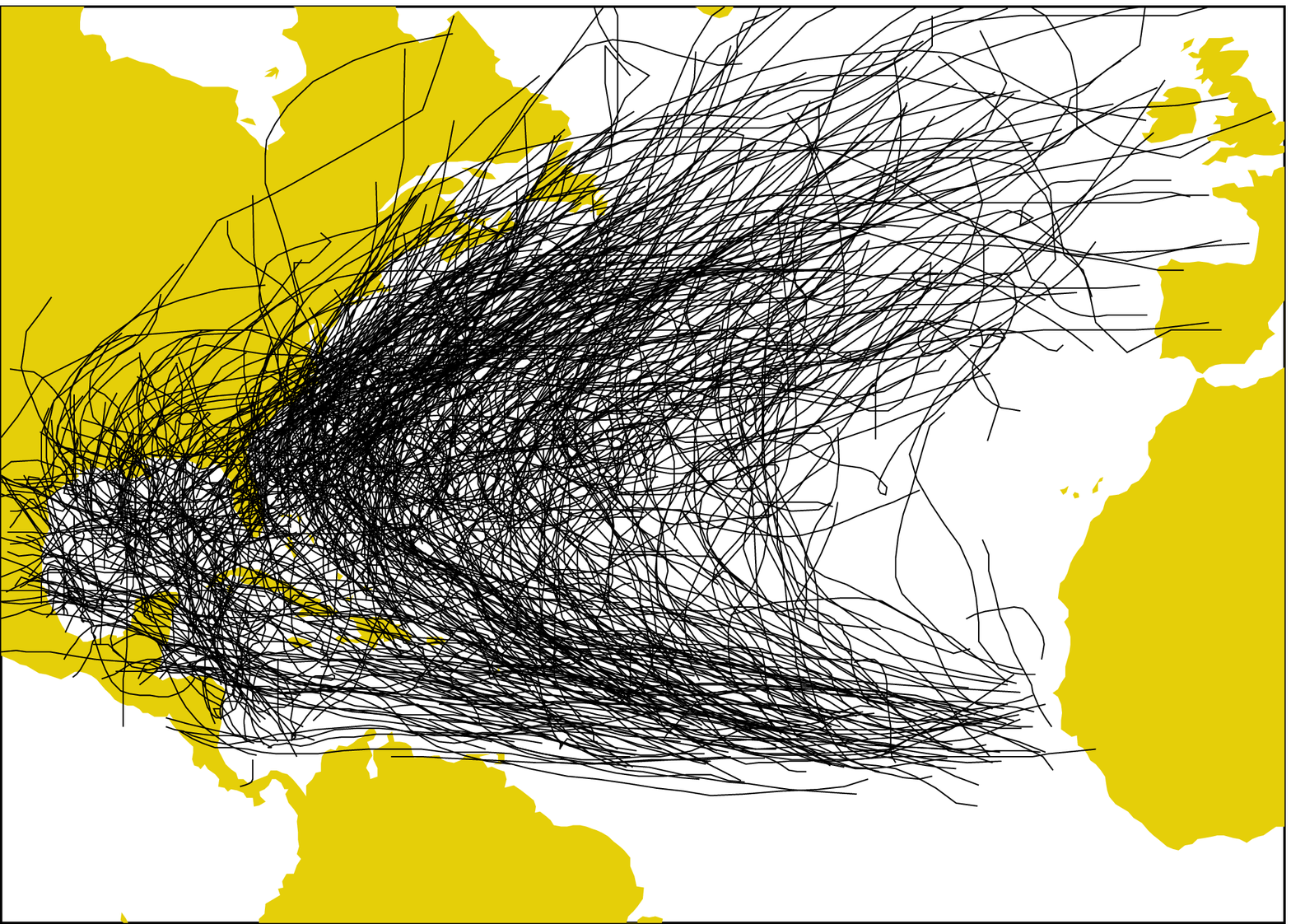}}
  \end{center}
  \caption{
Hurricane tracks from the HURDAT database for 1950 to 2003.
         }
  \label{f01}
\end{figure}

We use the HURDAT re-analysis `best track' data set of North
Atlantic tropical cyclones maintained by the NOAA's Hurricane
Research Division\footnote{\texttt{www.aoml.noaa.gov/hrd/hurdat}}.
HURDAT contains
6-hourly data of position of storm centers, as well as wind speed
and sea-surface pressure, for tropical cyclones from 1851 through
2003.  We restrict attention to dates after 1950, when Doppler
radar data started being used to make reliable wind-speed
estimates, and we remove a small fraction of storms that have
suspect features.  This leaves 524 storms, the data on which we
construct and evaluate our model. The tracks of these storms are
shown in figure~\ref{f01}.

\section{The model}

There are various statistical techniques available that one could
use to build a model of hurricane tracks. For instance, one could
consider parametric methods, non-parametric methods or
semi-parametric methods. For our purposes we will define these terms
as follows: in a parametric method, various parameters are estimated
from observational data. Simulations of hurricane tracks can then be
generated using these parameters without further recourse to the
data. In a non-parametric method, on the other hand, simulations of hurricane tracks
are generated directly from the observational data without any
parameters being estimated. Finally in a semi-parametric method
various parameters are estimated from the observational data, but
simulations are based on both the parameters \emph{and} the
observational data. We do not have any particularly strong views as
to which of these classes of techniques are generally more
appropriate for statistical modelling: for each particular problem
that one tackles one
should aim to derive models which combine simplicity with accuracy,
and depending on the nature of the problem such models may come from
any of these three classes. In this particular case
the model we use for the starting point for our hierarchy is
semi-parametric, which seems to offer a very
simple but nevertheless reasonable approach.

The two basic principles on which our model is based are that (a) the main
source of information we have about hurricane motion is the
observed motion of hurricanes in the historical record, and (b)
hurricanes in a similar part of the Atlantic tend to move in a
similar way. The second assumption is validated in our analysis. We
consider the motion of a hurricane over a 6 hour interval to consist
of a component along a line of longitude and a component along a
line of latitude. The size of each component is modelled as a sample
from a normal distribution, and the joint distribution of the two
components is taken as a bivariate normal distribution.
We do not argue that the bivariate normal distribution is the \emph{right}
distribution to model these displacements (i.e. we are not arguing that the
real displacements are sampled from a bivariate normal distribution), but
it certainly seems like a very reasonable choice for the first model in our
hierarchy. At a later stage we intend to test whether one can improve the results
by attempting to model various aspects of non-normality such as non-normal
marginals or a non-Gaussian copula.
To complete
the description of the model we simply need to specify the mean,
variance and correlation of the bivariate normal distribution. In this
article we will focus on modelling the unconditional mean displacement, which we define as
the expected displacement if we have no information about
previous motion of the hurricane, and know only its current location. In
subsequent studies we will consider models for the conditional mean
(conditioned on the track so far, and other factors) and the variance.

We will take the mean 6 hourly displacement of a hurricane currently at
the longitude-latitude point $(x,y)$ to be the weighted average of the 6 hourly displacements of all
hurricanes that have passed near the point $(x,y)$ in history. In
fact, we take the weighted average of the motion of \emph{all}
hurricanes, not only those that have passed near the point $(x,y)$,
but the weights will be negligible for hurricane tracks that are
far away. Our weighting model is based on a Gaussian,
with a length scale as a free parameter. We determine the optimal length
scale using a jackknife: we fit the model on $N-1$ years of data
and test it on the remaining year, and then repeat $N$ times
for the $N$ different possible choices of remaining year.

We can write our model for the unconditional mean displacement as:
\begin{equation}
 x_{new}=x_{old}+\Delta x
\end{equation}
where
\begin{equation}
E(\Delta x)=\frac{\sum \Delta x_i^h e^{-\frac{d^2}{\lambda^2}   }  }
                 {\sum e^{-\frac{d^2}{\lambda^2}}                  }
\end{equation}

where $x$ represent distance measured zonally, $x_{old}$ and $x_{new}$ are the zonal positions of a hurricane
at a 6 hourly interval,
 $\Delta x^h_i$ is a 6-hour displacement of a historical hurricane, $d$ is distance from $(x,y)$ to the
location of the historical displacement, $\lambda$ is a free parameter and the sums run over all
historical 6-hour displacements for the 524 hurricanes in our data set.
We use analogous expressions for displacements in the meridional direction.

What cost function should we use to fit the length-scale parameter $\lambda$?
Since we are ultimately interested in evaluating risks such as the risk of extreme winds in particular
locations, then one could argue that we should incorporate that directly into the cost function (e.g. the cost function could measure
the extent to which we simulate the observed pattern of extreme winds at selected locations during the last 50 years).
However, such a `macroscopic' cost function is rather hard to calculate, given that we are building up our
model `microscopically' from the motion of each hurricane track. And it may be flawed in that it may not
give unique solutions for the tracks, or it may be impractical to calculate in terms of CPU time, and thus
prohibit proper model testing.
However, this argument is moot anyway since the current model doesn't give predictions
of wind risk at all. So at this point we will use a `microscopic' cost function:
the RMSE of 6 hour forecast errors measured in km,
where the mean in the RMSE is taken over all observed 6 hourly displacements for each hurricane in the remaining year, and
over all years.
It is not entirely clear whether optimising using this microscopic cost function will be ultimately
sufficient for giving good predictions
of macroscopic quantities (although it seems likely that it might be), and we may have to revisit the question of
which cost function to use later once
the model is more complete.

Relative to the model of~\citet{emanuel05} (ERVR) we note that:
\begin{itemize}
    \item We have assumed bivariate normality for the distribution of 6 hourly
    track displacements, while ERVR avoid making any strong distributional assumptions
    \item We have avoided binning in space, while ERVR estimate transition probabilities on a grid
    \item We have avoided binning in speed or direction, while ERVR discretize the distribution
    of possible speeds and directions
\end{itemize}

Relative to the model of~\citet{vickery00} (VST) we note that:
\begin{itemize}
    \item Our model is semi-parametric (with one parameter), rather than parametric (with many parameters)
    \item We have assumed bivariate normality for the distribution of 6 hourly
    track displacements, while VST have assumed normally distributed errors on speed and direction
    \item We have avoided binning in space, while VST estimate parameters in 5$^{o}$x5$^{o}$ boxes.
\end{itemize}

We have designed our model as described above because we think this is a sensible, simple and promising way to build such a model.
However, we do not argue that our model is ultimately \emph{better} than these other models. That is not for us to
decide, but would be a question for statistical testing. For instance, either of the above models could be used to make a
cross-validation estimate of our RMSE-based cost function.

\section{Results}

\begin{figure}[!htb]
  \begin{center}
    \scalebox{0.8}{\includegraphics{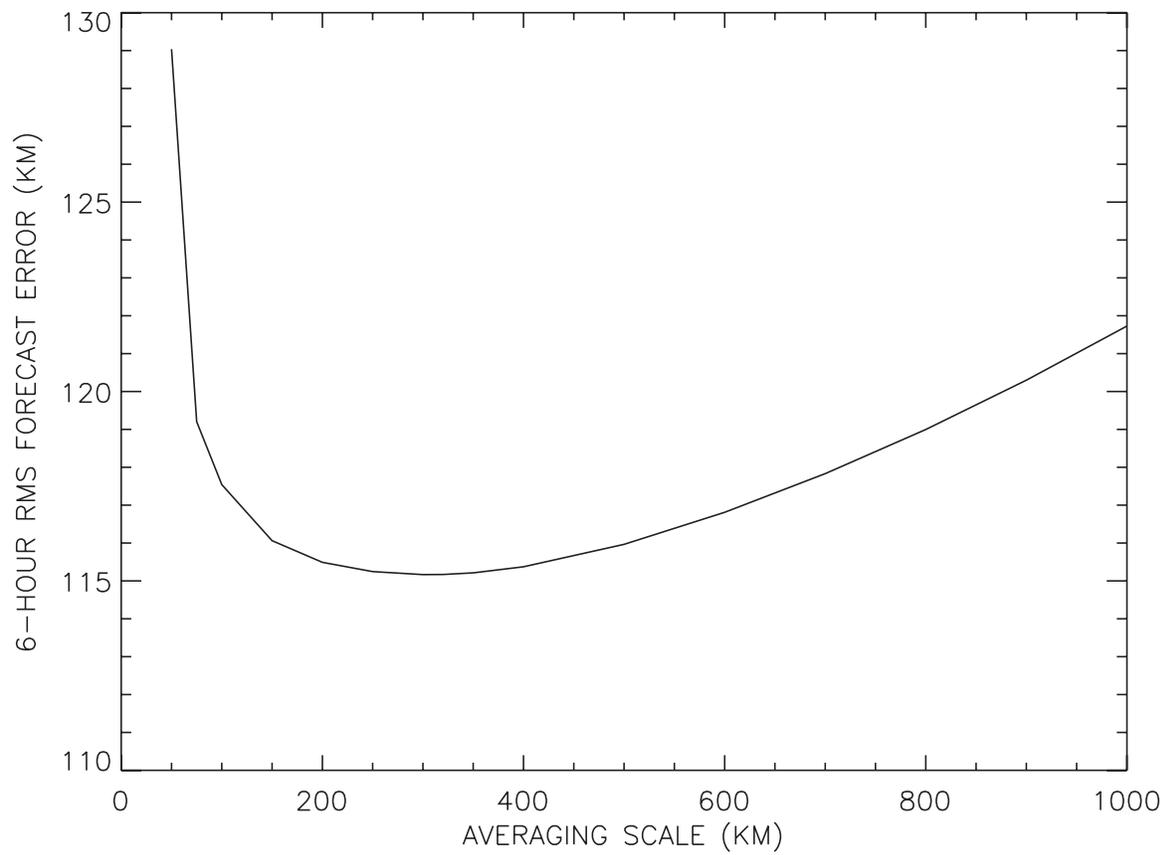}}
  \end{center}
  \caption{
The cost function for the fitting of the hurricane track model described in the text,
calculated using a jackknife.
         }
  \label{f02}
\end{figure}

We can test our model for any value of the length-scale $\lambda$, and can calculate the value for the
cost function in each case. The results of this exercise are shown in figure~\ref{f02}.
We see that for very short length-scales the model does badly (gives large forecast errors)
presumably because not enough data is
being used to be able to estimate the likely hurricane displacements in an accurate way.
Another way to say the same thing would be that
the results are too affected by sampling error due to the effects of individual synoptic systems on the tracks
of individual hurricanes.
For very long length-scales the model also does not perform particularly well, presumably because the
information being used includes information from far away, in regions where hurricanes move
differently due to different climatological conditions.

The optimum value for the length-scale is around 300km, although the minimum is rather broad,
especially on the higher length-scale side.
This broad minimum seems to imply that we have more than enough data to determine the unconditional
mean hurricane motion fairly accurately.
This is presumably because there is a high density
of data relative to the scales inherent in the shapes of the mean hurricane tracks, and so the variations
in the shape of the mean hurricane tracks are well sampled. This is encouraging, and makes us think
that our model is well fitted. It also suggests that there may be enough data to try and
condition the shapes of the tracks on various covariates such as time of year, state of ENSO, state of NAO,
intensity of the hurricane, etc. We will look at this in a future study. The risk of such conditioning is,
of course, that one might overfit the model, and one would need to use appropriate tests to avoid that.
By contrast, had the data been sparse,
then the inclusion or exclusion of individual tracks by changes in
the averaging length-scale would strongly affect the mean track and
forecast error, leading to more structure in the forecast error
function.

\begin{figure}[!htb]
  \begin{center}
    \scalebox{0.8}{\includegraphics{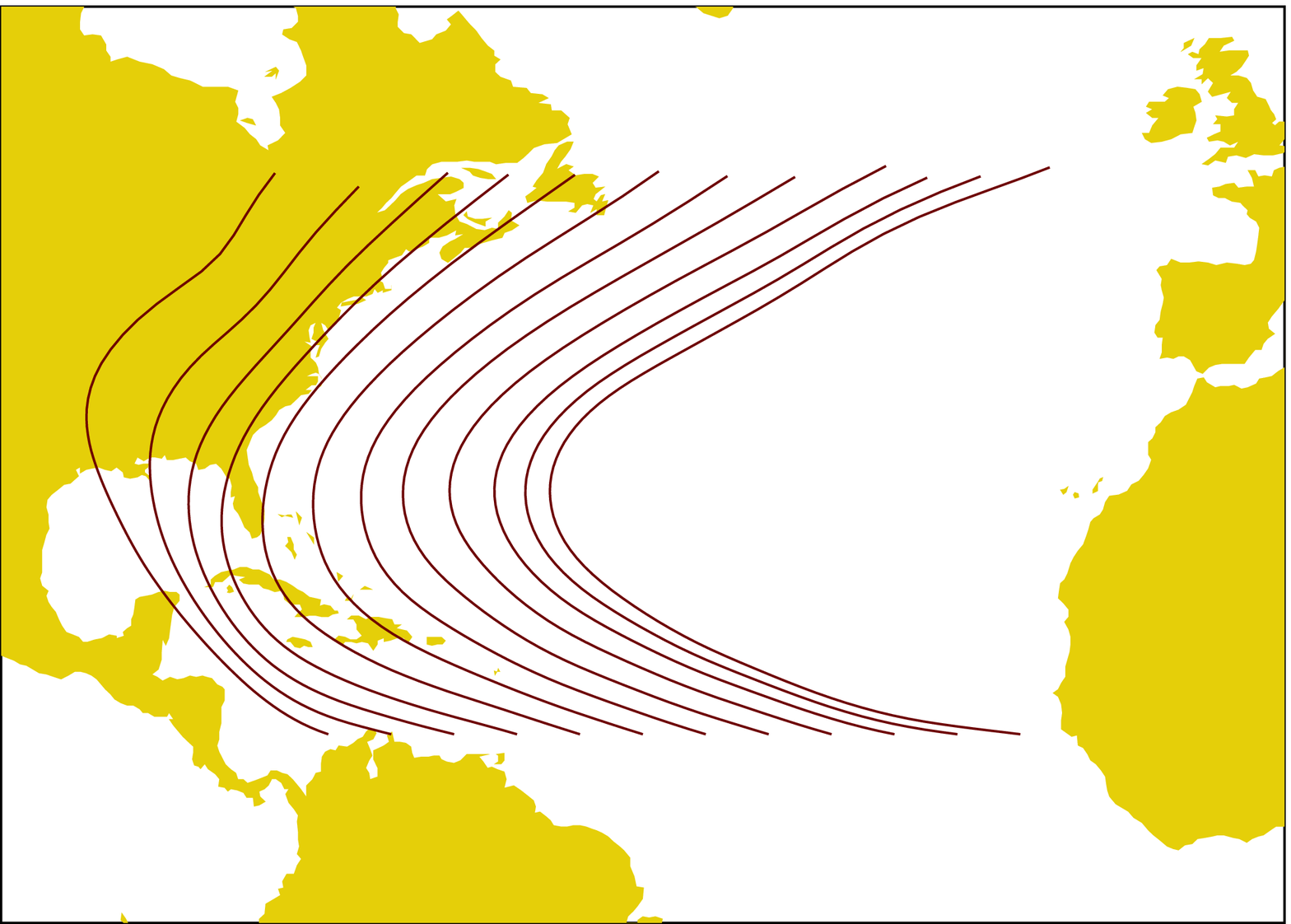}}
  \end{center}
  \caption{
Estimated unconditional mean hurricane tracks, based on the optimal length-scale
of 300km.
         }
  \label{f03}
\end{figure}

Having derived the optimum length scale we can calculate the mean hurricane track displacements
at any point in the basin, and can integrate along these displacements to form `mean hurricane tracks'.
The results of this are shown in figure~\ref{f03}. We have integrated these tracks from a line of points
at 10$^{o}$N, spaced at intervals of 10$^{o}$.
Because we are only modelling the unconditional mean motion of hurricanes, and not the variance, our tracks are entirely
deterministic and don't cross each other.

Figure~\ref{f03} shows the characteristic curving of hurricane tracks, and little else.
This characteristic shape
can be explained as due to the local Hadley circulation, consisting of low pressure near the equator and
high pressure near the tropic of capricorn. The mid and lower level tropospheric winds in this circulation
are responsible for steering hurricanes, and the average circulation of these winds creates the
curving shape of the mean hurricane tracks.

\begin{figure}[!htb]
  \begin{center}
    \scalebox{0.8}{\includegraphics{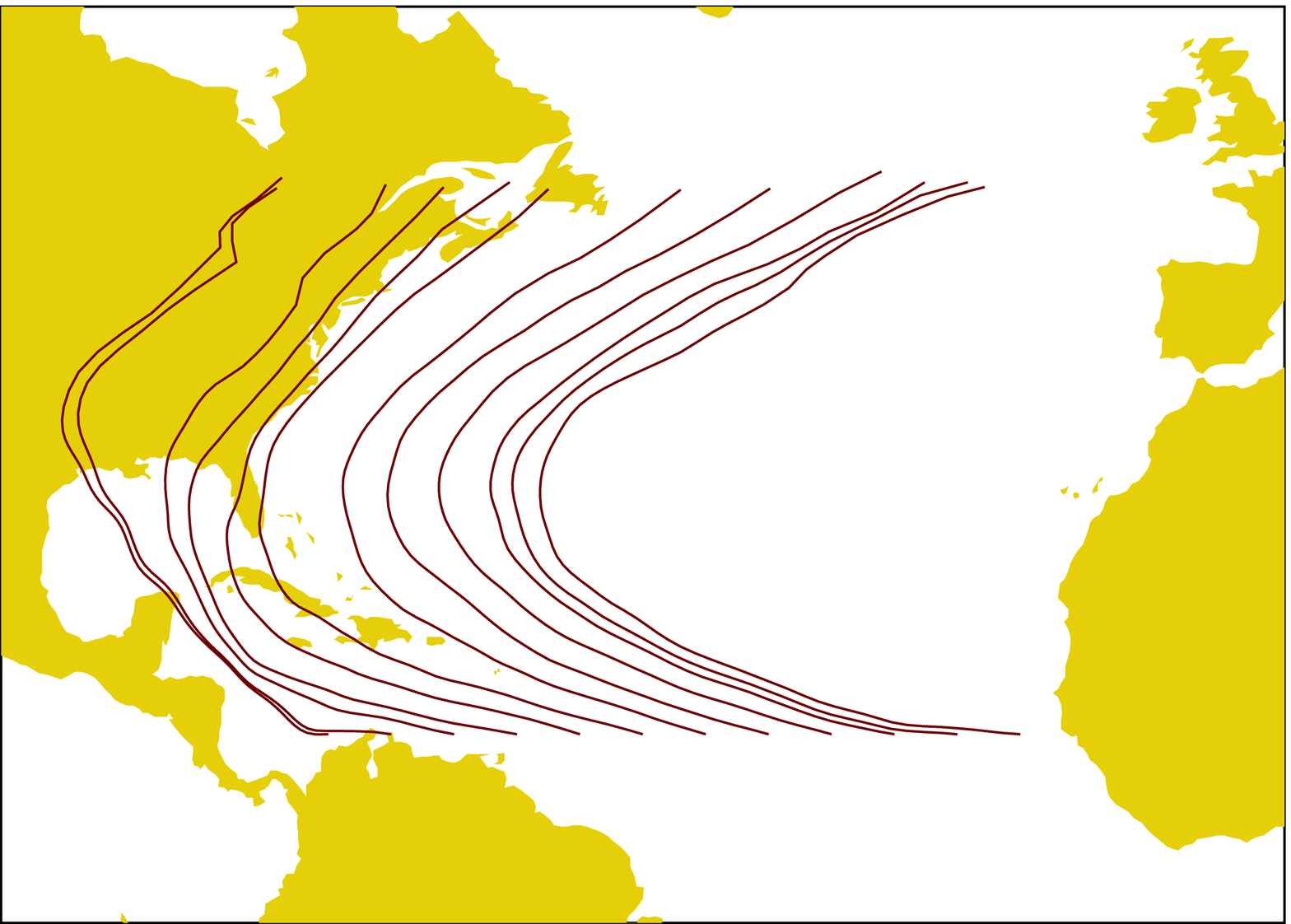}}
  \end{center}
  \caption{
Estimated unconditional mean hurricane tracks, based on a length scale of 100km,
which is shorter than the optimal length scale of 300km.
         }
  \label{f04}
\end{figure}

What if we were to use a length scale that is much shorter than the optimum?
Results from using a length-scale of only 100km are shown in figure~\ref{f04}.
We see that the modelled mean tracks are less smooth, presumably because they
are now too closely constrained to individual historical storm tracks.

\begin{figure}[!htb]
  \begin{center}
    \scalebox{0.8}{\includegraphics{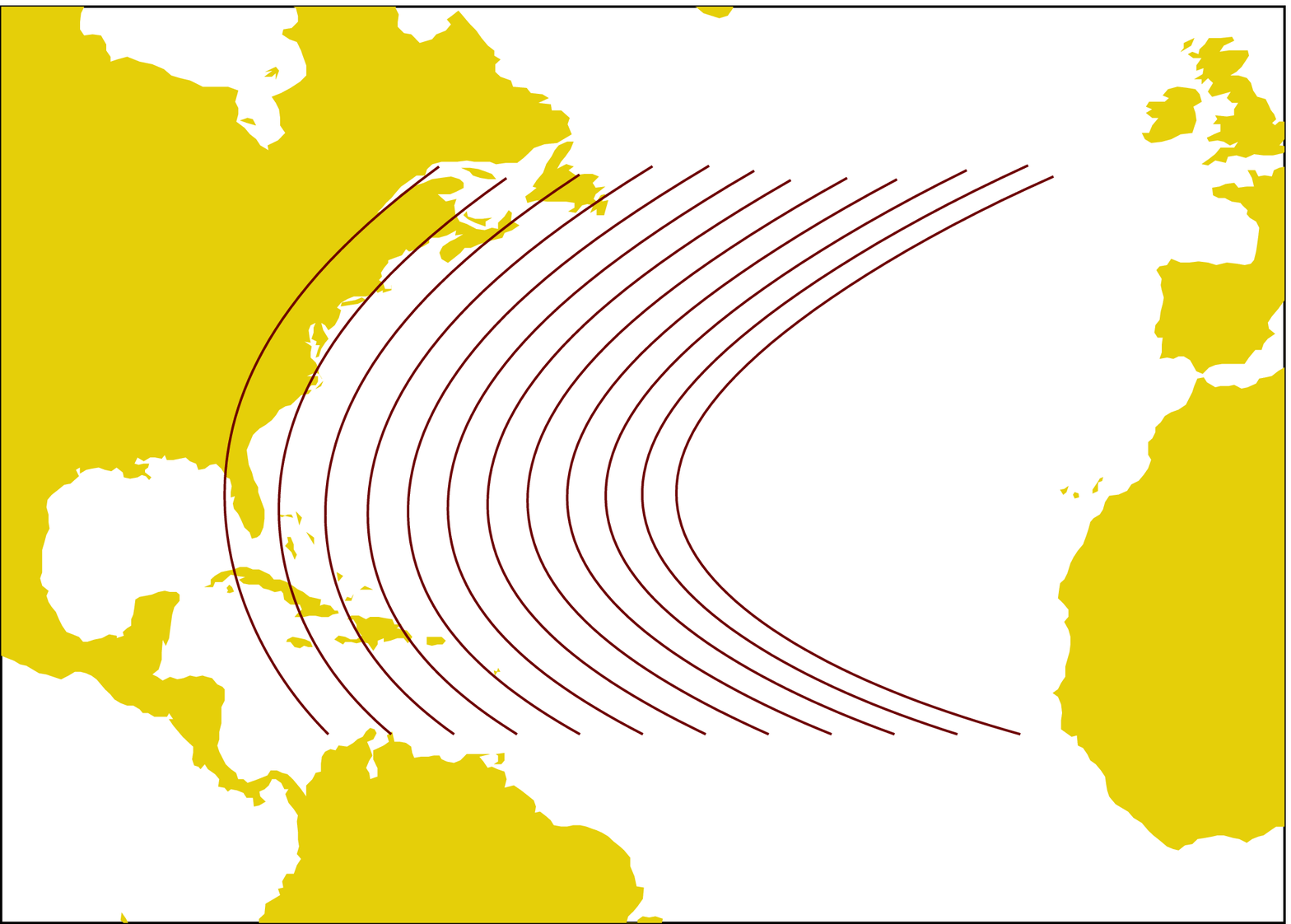}}
  \end{center}
  \caption{
Estimated unconditional mean hurricane tracks, based on a length scale of 1000km,
which is longer than the optimal length scale of 300km.
         }
  \label{f05}
\end{figure}

What if we were to use a length scale that is much \emph{longer} than the optimum?
Results from using a length-scale of 1000km are shown in figure~\ref{f05}.
We see that the characteristic
curving shape of hurricane tracks is rather smoothed out.

\begin{figure}[!htb]
  \begin{center}
    \scalebox{0.8}{\includegraphics{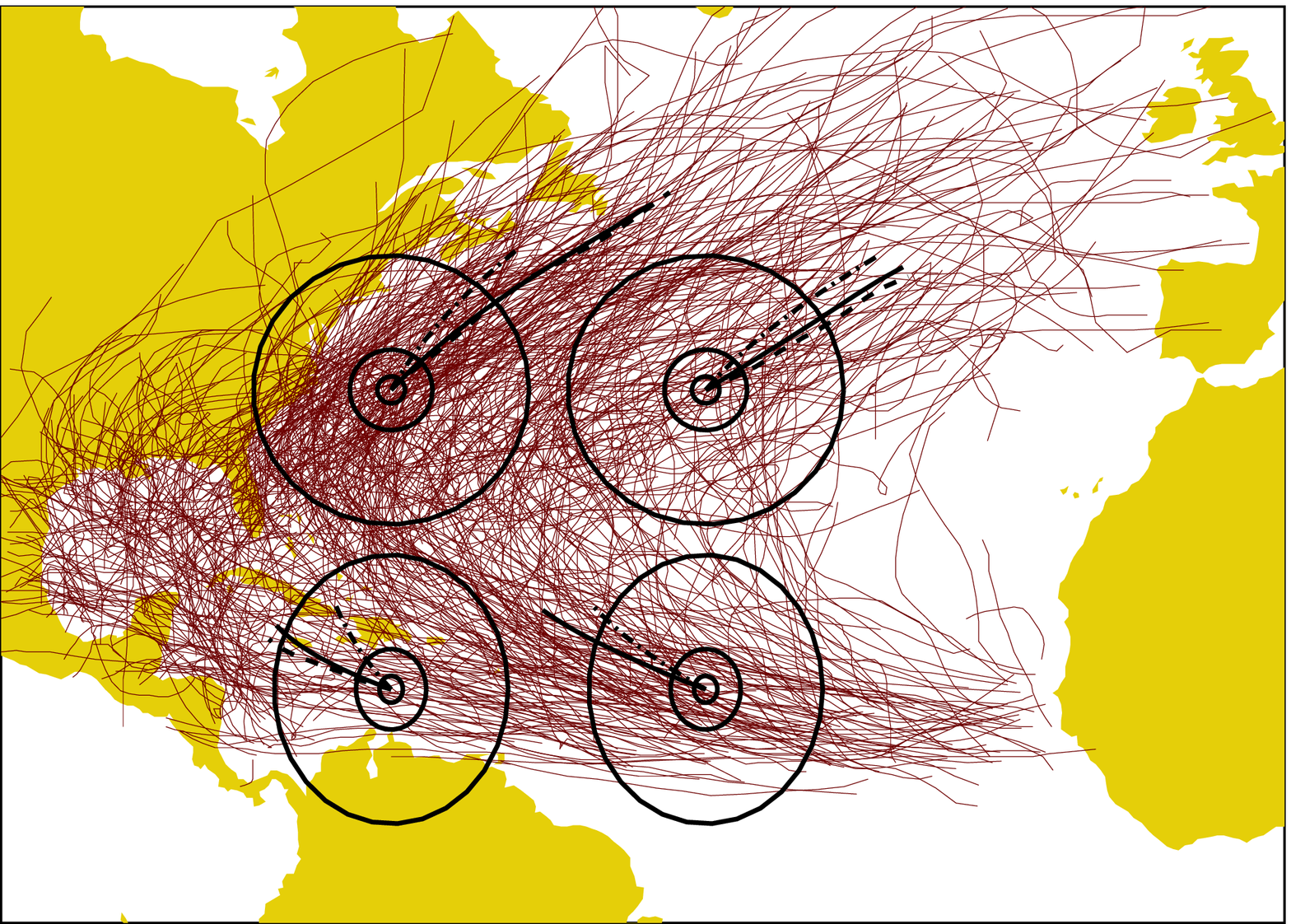}}
  \end{center}
  \caption{
Historical hurricane tracks (as shown in figure~\ref{f01}) with circles
overlaid with radii of 100km, 300km and 1000km. 300km is the optimal
smoothing length-scale for the model described in the text. The black tracks
emerging from the circles are mean hurricane tracks derived using length-scales
of 100km (dashed line), 300km (solid line) and 1000km (dot-dashed line).
         }
  \label{f06}
\end{figure}

Figure~\ref{f06} shows historical hurricane tracks (exactly as in figure~\ref{f01})
but with circles of radii 100km, 300km and 1000km at four points in the North Atlantic,
to illustrate the size of the regions over which hurricane tracks are being averaged in the model.
It also shows unconditional mean tracks derived using these three length-scales at each
of these points (dashed lines for the 100km length-scale, solid lines for 300km and dot-dashed lines for 1000km).
It seems to be possible to understand the differences between the mean
tracks from the three different length-scales in some of the cases shown in figure~\ref{f06}.
For instance, for the most southerly and westerly
of the four points, the mean track derived from the 1000km length-scale lies to the right
of the mean track derived from the shorter length-scales. This is perhaps because averaging using
the 1000km length-scale includes tracks from north of Hispaniola that are curving
northward more quickly than those from south of Hispaniola.

\begin{figure}[!htb]
  \begin{center}
    \scalebox{0.8}{\includegraphics{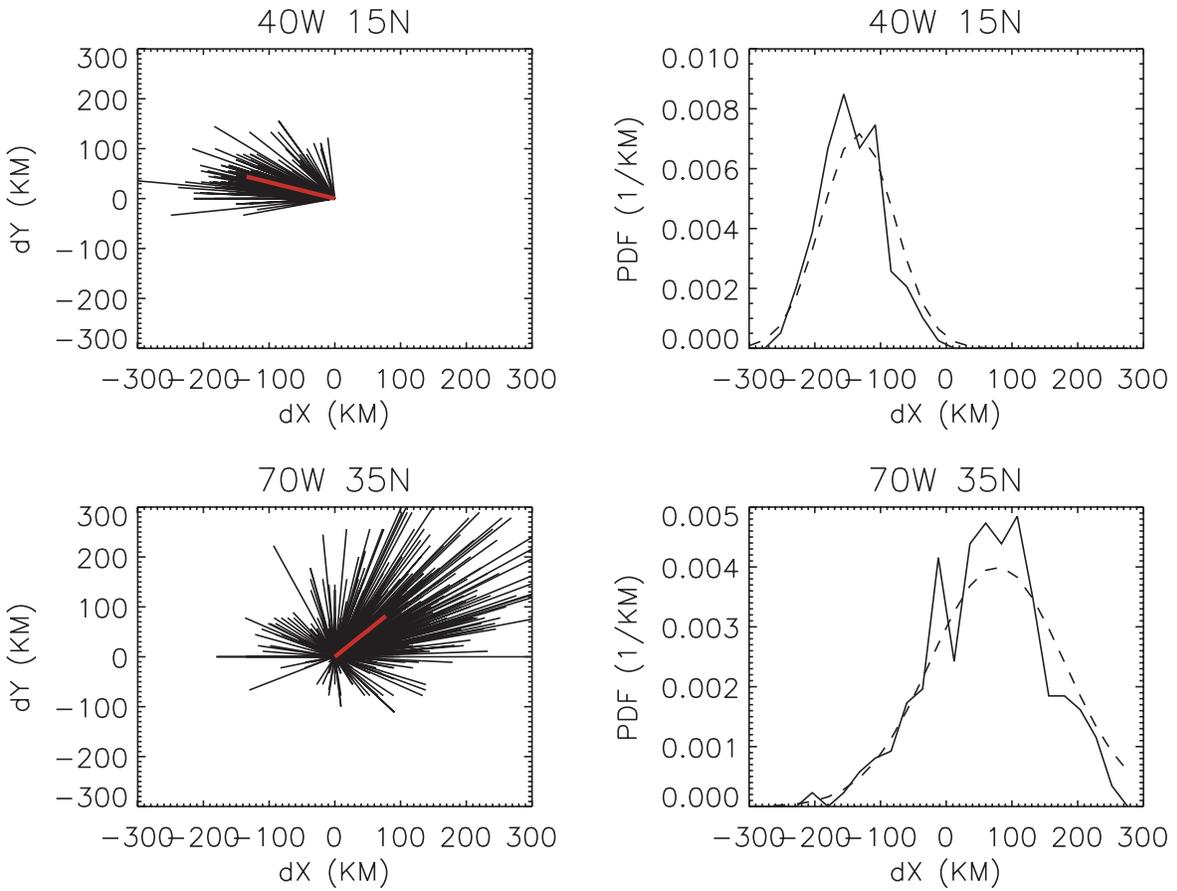}}
  \end{center}
  \caption{
In the left column, vectors showing historical hurricane displacements within
300km of two points in the North Atlantic (black lines), along with a mean vector
derived by averaging the zonal and meridional components of all the individual vectors.
In the right column, the estimated
distribution of the zonal displacements from these vectors (solid line) along
with a fitted normal distribution (dashed line).
         }
  \label{f07}
\end{figure}

Finally, in figure~\ref{f07}, we show some diagnostics which give further insight
into how the track model is working. The left hand panels show the displacement vectors
from the historical record that lie within 300km of two selected locations in the
North Atlantic. The red vector shows the vector generated by averaging the zonal and
meridional components of these vectors separately (this averaging is not quite the same
as the averaging process used in the model described above, but is very similar).
The right hand panels show the distribution of zonal displacements for these vectors
(solid line) along with a fitted normal distribution (dashed line). In these two cases
we see that a normal distribution is not a bad model. In other locations the normal
doesn't perform as well, and testing more complex distributional models is a priority
for further investigation.

\section{Discussion}

We have described the results of our first steps towards building a statistical model of hurricane tracks.
At this point we have only considered how to model unconditional mean hurricane tracks, and we have not considered
fluctuations around these mean tracks that would be caused by the synoptic situation of the day.
We have shown how these mean tracks can be derived very simply using a simple semi-parametric model with
just a single parameter. We calculate the optimal value for this parameter using a jackknife fitting procedure.

The method we use to derive the mean tracks seems to be a reasonably good scheme for summarising the
mean behaviour of hurricane tracks, and we plan to apply it to various subsets of the full track set in order to
investigate the effects of various conditioning factors such as time of year, intensity of the storm,
state of ENSO and state of the NAO.

It may be possible to improve on our simple model for the unconditional mean tracks, perhaps
by conditioning on time of year or intensity, or maybe by changing the precise form of the weighting
function (i{.}e{.} by using $e^{-\frac{d}{\lambda}}$, a top-hat, or similar). We plan to test this.

The next stage in the development of our model hierarchy will be
modelling the fluctuations around the mean tracks.
We also plan to apply our model to tropical storm tracks in other basins, and to extratropical
storm tracks.

\section{Legal statement}

SJ was employed by RMS at the time that this article was written.

However, neither the research behind this article nor the writing
of this article were in the course of his employment, (where 'in
the course of their employment' is within the meaning of the
Copyright, Designs and Patents Act 1988, Section 11), nor were
they in the course of his normal duties, or in the course of
duties falling outside his normal duties but specifically assigned
to him (where 'in the course of his normal duties' and 'in the
course of duties falling outside his normal duties' are within the
meanings of the Patents Act 1977, Section 39). Furthermore the
article does not contain any proprietary information or trade
secrets of RMS. As a result, the authors are the owners of all the
intellectual property rights (including, but not limited to,
copyright, moral rights, design rights and rights to inventions)
associated with and arising from this article. The authors reserve
all these rights. No-one may reproduce, store or transmit, in any
form or by any means, any part of this article without the
authors' prior written permission. The moral rights of the authors
have been asserted.

The contents of this article reflect the authors' personal
opinions at the point in time at which this article was submitted
for publication. However, by the very nature of ongoing research,
they do not necessarily reflect the authors' current opinions. In
addition, they do not necessarily reflect the opinions of the
authors' employers.

\bibliography{timhall2}

\clearpage

\clearpage

\clearpage

\clearpage

\clearpage

\clearpage

\clearpage

\end{document}